\documentclass[useAMS,usenatbib]{mn2e}
\usepackage{graphicx}
\usepackage{epsfig}
\usepackage{natbib}
\usepackage{lscape}
\usepackage{subfigure}
\usepackage{verbatim}
\usepackage{amssymb,amsmath}
\newcommand{\prate}{\dot\chi}
\DeclareMathVersion{bold}

\title[The HTRU Survey Single-Pulse Search]{The High Time Resolution Universe Survey -- III. Single-pulse searches and preliminary analysis}

\author[S. Burke-Spolaor et al.]{ 
S. Burke-Spolaor,$^{1,2}$\thanks{Email: sarah.spolaor@csiro.au}
M. Bailes$^{1}$,
S. Johnston$^{2}$,
S.~D. Bates$^{3}$,
N.~D.~R. Bhat$^{1}$,\newauthor
M. Burgay$^{4}$,
N. D'Amico$^{4}$,
A. Jameson$^{1}$,
M.~J. Keith$^{2}$,
M. Kramer$^{3,5}$,
L. Levin$^{1,2}$,\newauthor
S. Milia$^{4,6}$,
A. Possenti$^{4}$,
B. Stappers$^{3}$,
W. van Straten$^{1}$\\
$^{1}$Centre for Astrophysics and Supercomputing, Swinburne University of Technology, PO Box 218 Hawthorn, VIC 3122, Australia\\
$^{2}$Australia Telescope National Facility, CSIRO, P.O. Box 76, Epping, NSW 1710, Australia\\
$^{3}$Jodrell Bank Ctr. for Astrophysics, School of Physics and Astronomy, University of Manchester, Manchester M13 9PL, UK\\
$^{4}$INAF - Osservatorio Astronomico di Cagliari, Poggio dei Pini, 09012 Caopterra, Italy\\
$^{5}$MPI fuer Radioastronomie, Auf dem Huegel 69, 53121 Bonn, Germany\\
$^{6}$Dipartimento di Fisica, Universit\`a degli Studi di Cagliari, Cittadella Universitaria, 09042 Monserrato (CA), Italy
}

\date{}
\begin{document} 

\maketitle
\begin{abstract}
  We present the search methods and initial results for transient radio
  signals in the High Time Resolution Universe (HTRU) Survey. The HTRU
  survey's single-pulse search, the software designed to perform the search,
  and a determination of the HTRU survey's sensitivity to single pulses are
  described. Initial processing of a small fraction of the survey has produced
  11 discoveries, all of which are sparsely-emitting neutron stars, as well as
  provided confirmation of two previously unconfirmed neutron stars. Most of
  the newly discovered objects lie in regions surveyed previously, indicating
  both the improved sensitivity of the HTRU survey observing system and the
  dynamic nature of the radio sky. The cycles of active and null states in
  nulling pulsars, rotating radio transients (RRATs), and long-term
  intermittent pulsars are explored in the context of determining the
  relationship between these populations, and of the sensitivity of a search
  to the various radio-intermittent neutron star populations. This analysis
  supports the case that many RRATs are in fact high-null-fraction pulsars
  (i.\,e.~with null fraction $\gtrsim$0.95), and indicates that intermittent
  pulsars appear distinct from nulling pulsars in their activity cycle
  timescales. We find that in the measured population, there is a deficit of
  pulsars with typical emission timescales greater than $\sim$300\,s that is
  not readily explained by selection effects. The HTRU low-latitude survey
  will be capable of addressing whether this deficit is physical. We predict
  that the HTRU survey will explore pulsars with a broad range of nulling
  fractions (up to and beyond $0.999$), and at its completion is likely to
  increase the currently known RRATs by a factor of more than two.
\end{abstract}
\begin{keywords}
\end{keywords}

\section{Introduction}\label{sec:htruintro}
Radio single-pulse searches within the last decade have revealed a number of
fascinating objects within and outside of our own Galaxy. Pulsed radio
emission originating from other galaxies could give an estimate of the
combined free electron content of their host galaxies and of the intergalactic
medium; thus far, searches have detected single pulses from neutron stars in
two close satellites of our Galaxy, including giant pulses from B0540--69 in
the Large Magellanic Cloud \citep{johnstonromani03} and a probable transient
neutron star in the Sculptor Spheroidal Dwarf Galaxy \citep{herrerathesis}. A
number of other nearby galaxies have been the target of single-pulse searches,
showing only sparsely detected events of inconclusive origin (e.\,g.\,galaxies
M33 and M31, searched by \citealt{mclaughlincordes03} and
\citealt{herrerathesis}, respectively). The discovery by \citet{LB} of a
seemingly dispersed, 30-Jy impulse (at a dispersion measure, DM, of
$375$\,pc\,cm$^{-3}$) appeared to represent the first discovery of an
extragalactic pulse with high significance that was of non-neutron star
origin. However, the recent discovery of pulses of ambiguous terrestrial
origin with frequency-sweeps that mimic the cold plasma dispersion relation
and primarily appear around $DM\simeq370$\,pc\,cm$^{-3}$, have cast some doubt
on the Lorimer et al.~pulse as being extragalactic and provided an additional,
terrestrial target for single-pulse searches \citep{perytons}.

The most frequently detected sources of transient radio emission are
pulsars. Single-pulse searches in recent years uncovered what was labelled the
``rotating radio transient'' (RRAT) phenomenon \citep{rrats}, which
encapsulates sparsely radio-emitting neutron stars, that due to their sporadic
emission, were undetectable by the Fourier-based search techniques currently in
standard use for pulsar surveys. The point has recently been made that in fact
the term RRAT is rather a ``detection label'' than a description of a physical
phenomenon: Keane (2010a)\nocite{keaneproceeding} notes that whether a
discovery is considered ``RRAT'' is highly dependent on both survey length and
an object's rotational period. This and other mounting evidence indicate that
RRATs are unlikely to have a physically distinct origin from radio pulsars.
This includes, for instance, that RRAT pulse energy distributions tend to obey
log-normal distributions similar to both non-nulling and nulling pulsars
(e.\,g.~\citealt{cjd01}, Keane 2010b\nocite{evanthesis}; Miller et
al.~\emph{submitted}\nocite{miller}) and that there is a lack of distinction
between RRAT and pulsar Galactic distributions and pulse width distributions
\citep{sbsmb}. Additionally, if RRATs were a distinct phenomenon unrelated to
other neutron star populations in an evolutionary sequence, the implied
birthrate of neutron stars would far exceed the rate of supernovae that
produce them \citep{evan}.  As such, the current understanding of the objects
discovered preferentially in single-pulse searches is that they are a mix of
both modulated pulsars with long-tailed pulse energy distributions
\citep{modpulsar}, and pulsars at the most extreme end of the nulling pulsar
population (\citealt{sbsmb}; Keane 2010b\nocite{evanthesis}; Miller et
al.~\emph{submitted}).

However, questions remain: why do the objects appear to have a
period-derivative and/or magnetic field distribution that sits higher than
average pulsars of the same period range \citep{mauranewpaper}? What are the
statistics of peculiar phenomena in these extreme nulling objects such as the
RRAT-pulsar mode switching of PSR J0941--39 \citep{sbsmb}, the multi-modal
(and latitude-dependent) behaviour of PSR J1119--6127 \citep{trinity}, or the
glitch activity of PSR J1819--1458 \citep{lyne1819}?  What causes nulling and
what is the distribution of pulsar nulling fractions (again holding
implications for the neutron star birthrate)? And finally, do (and if so, how
do) the most extreme nulling pulsars fit into an evolutionary progression
between average radio pulsars, nulling pulsars, radio quiet neutron stars, and
the magnetar population (e.\,g.~\citealt{mauranewpaper,lyne1819,sbsmb}; Keane
2010b\nocite{evanthesis})?
Acquiring larger statistical sample of these objects is among the next
essential steps in understanding RRATs and transient radio neutron star
phenomena (currently a total of $\sim$40 such objects are known,
\citealt{hesselsproc,deneva,keane,sbsmb,herrerathesis}).

The High Time Resolution Universe (HTRU) Survey is the first digital,
all-southern-sky survey for pulsars and fast (sub-second) transients, covering
declinations \mbox{$\delta<+10^\circ$} (\citealt{HTRU1}; hereafter HTRU Paper
1).  The survey employs a new digital backend, the ``Berkeley-Parkes-Swinburne
Recorder'' (hereafter BPSR), that has been installed at the Parkes Telescope
for the 20-cm multibeam receiver \citep{multibeam}. BPSR allows improved
digitisation levels, frequency and time resolution over the previous analogue
instrument that has been used for previous southern single-pulse searches and
studies (i.\,e.~\citealt{sbsmb},\citet{rrats},\citet{keane}).
This affords the HTRU survey unprecedented sensitivity to sub-second duration,
dispersed single impulses of radio emission in the southern sky. This paper
describes the techniques used in the survey to search for single pulses.

As an all-sky survey, the HTRU survey is sensitive to terrestrial, Galactic,
and extragalactic sources of radio pulses. The HTRU intermediate and high
latitude surveys will cover Galactic longitudes $-120^\circ<l<30^\circ$ and
latitudes $|b|<15^\circ$, and declinations $\delta<+10^\circ$ (not covered by
intermediate pointings), respectively. Although no processing has yet
commenced for the HTRU low-latitude survey, that survey will cover areas
$|b|<3.5^\circ$ and $-80^\circ<l<30^\circ$, with 70\,minutes per pointing. A
parallel northern effort is underway at Effelsberg Radio Telescope, affording
a full sky survey for pulsars and transients. The low and intermediate
latitude surveys share common areas of sky coverage to previous surveys which
employed the analogue filterbank installed on the Parkes 20-cm Multibeam
receiver. The ``Parkes Multibeam Survey'' \citep{pksmb} covered $|b|<5^\circ$
with 35-minute pointings, and was searched for single pulses by \citet{rrats}
and \citet{keane}, revealing the bulk of the currently recognised RRATs.
Hereafter we refer to those two searches collectively as the PKSMB searches.
At higher Galactic latitudes, a single-pulse search was performed on the
\citet{ED} and \citet{BJ} surveys by \citet{sbsmb} (hereafter BSB). Despite
the common sky coverage and high success rate of the BSB and PKSMB searches,
we nevertheless expect to discover new sources of pulsed radio emission in
these areas; some pulsing radio sources may not have been previously detected
due to the source's low event rate, insufficient sensitivity in previous
surveys, or the improved dynamic range of the BPSR observing system over the
previous analogue filterbank system installed on the Parkes multibeam
receiver, whose dynamic range was hurt by $\sqrt{N}$-statistics (explored in
more detail in \S\ref{sec:htrusens} below).


In this manuscript, we present the search methods, sensitivity, and current
status of the HTRU survey single-pulse search
(\S\ref{sec:htrusearch}). Because this survey required concurrent searches
using periodicity and single-pulse techniques, it affords the opportunity to
explore efficient modes of search operation, attempting to perform these
techniques in parallel with a minimum of duplicated computing effort. We also
report on initial discoveries in the intermediate- and high-latitude surveys
(\S\ref{sec:htrunewdets}). In \S\ref{sec:htrudiscussion}, we explore the
detectability of our discoveries in previous surveys of overlapping regions,
present our new discoveries in terms of the windowing description of pulsar
intermittence presented in BSB, and make predictions for the sparsely-emitting
radio neutron stars that the full HTRU survey will uncover.

\section{The HTRU Survey's Single-pulse (SP) Search}\label{sec:htrusearch}

\begin{figure*}
\centering
\includegraphics[angle=270,width=0.85\textwidth,trim=3mm 5mm 12mm 50mm, clip]{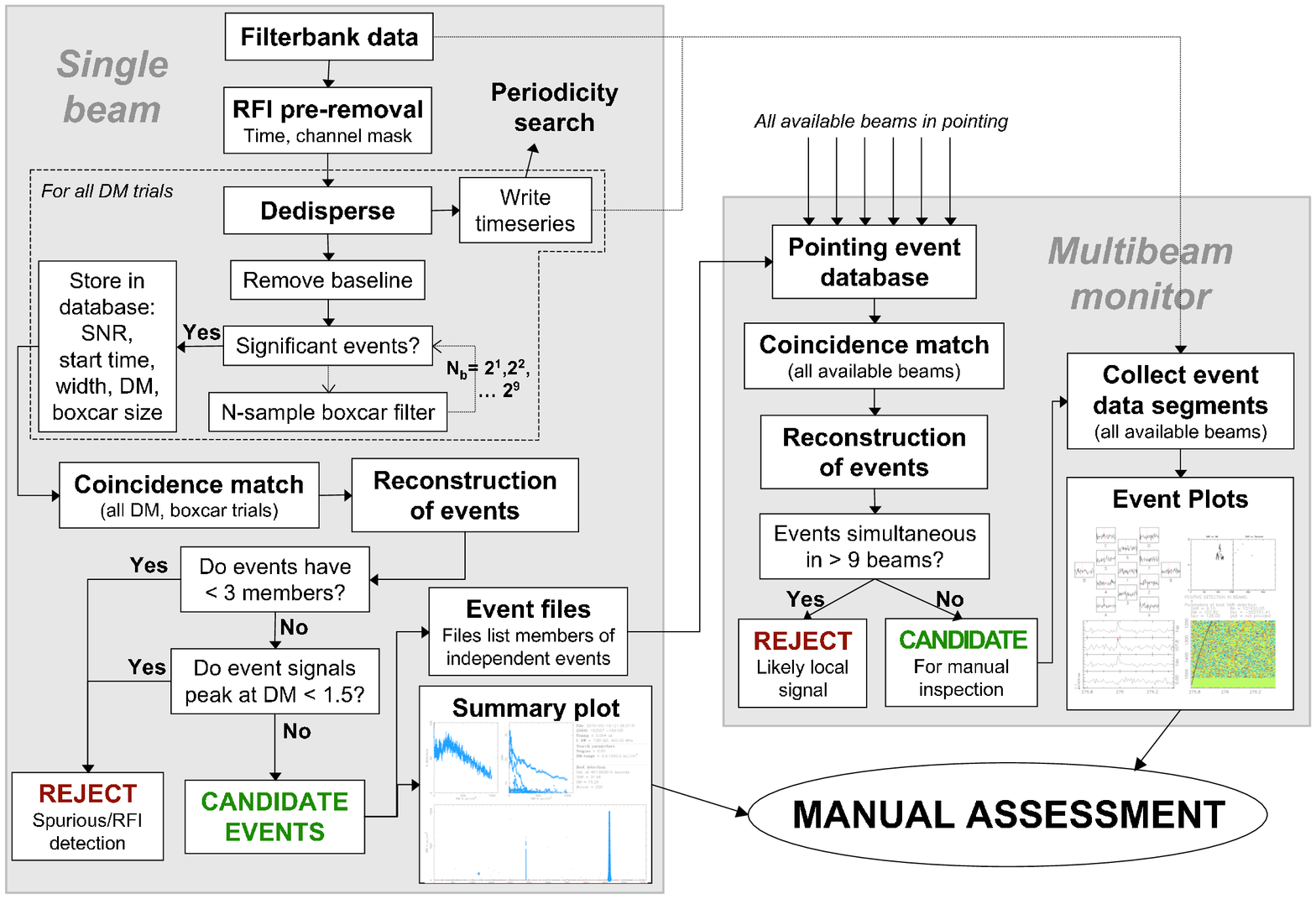}
\caption[The HTRU single pulse pipeline]{A diagrammatic representation of the
  HTRU survey single-pulse pipeline. Each of up to 13 ``single beam''
  processes associated with one multibeam pointing are run simultaneously,
  distributed onto several compute nodes and producing single-pulse and
  periodicity search results for that beam. The multibeam monitor tracks the
  status of the individual beams, collecting single-pulse search results,
  performing interference excision based on multiple-beam occurrence, and
  producing single-event diagnostic plots for each event in the pointing not
  tagged as interference.}\label{fig:sppipe}
\end{figure*}

\subsection{Data Analysis Pipeline}\label{sec:htrusppipe}
Thus far all processing and analysis has been carried out using the ``Green
Machine'' supercomputer at the Swinburne Centre for Astrophysics and
Supercomputing. The pipeline employs several standard search techniques
(e.\,g.~dedispersion and boxcar matched filtering, multi-beam coincidence
matching; \citealt{cordesmclaughlin03,deneva,keane}; BSB), and several novel
techniques (e.\,g.~independent-event identification and inspection). The basic
methodology of the SP search is based on that described by BSB.

There are two main bottlenecks in the HTRU survey's single-pulse search. The
first is the computational load demanded by the dedispersion of our data, and
the second is the number of candidates which need to be manually assessed. The
single-pulse data reduction pipeline design includes schema that aim to abate
both of these. First, we minimise the net computational time of the
periodicity and SP searches by integrating the SP search into the {\sc hitrun}
processing pipeline introduced in HTRU Paper 1. By doing this, the
interference excision and dedispersion described in HTRU Paper 1 is done only
once, producing each data stream's results simultaneously for use in the
single-pulse search and the periodicity search. As described below, the time
series search for single pulses is performed while each time series is still
in the computer's ``random access memory'', and for each 9-minute data stream
adds $<$2.5 minutes of processing time onto the total $\sim$3-hour timescale
for the {\sc hitrun} Fourier-based pipeline. We stress that the added
computational load of including a single pulse search to a traditional Fourier
search is negligible.

We address the issue of large candidate numbers by identifying independent
events across multiple dispersion, boxcar trials, and beams, and by performing
a simple threshold-based interference zapping.
The simultaneous processing and viewing of data from multiple beams decreases
the occurrence of candidates caused by low-level, local interference and
greatly increases the speed at which the manual assessment of a candidate can
be made. However, due to limited computational resources, we cannot store all
13 beams on one node, requiring the use of a monitor to track and collect
beams from each pointing as they are processed.

The single-beam SP search and the multibeam data monitor are described below.
Figure \ref{fig:sppipe} gives a schematic diagram for the pipeline, showing
one beam and the multibeam monitor.

\subsubsection{Single-beam Processing}\label{sec:singlebeam}
For each observation, a multibeam monitor is initiated and the available beams
for that pointing are submitted in sequence for processing on the
supercomputer, each processed independently using the {\sc hitrun} pipeline.
For this pipeline we begin with 2-bit filterbank data with the sampling
properties of which are summarised in Table \ref{table:parameters}.
Pre-analysis radio frequency interference excision is done first, as detailed
in HTRU Paper 1. In brief, this excision creates and uses both a frequency and
a time mask that flags periodic-interference-affected channels and time
samples with a ${\rm DM}=0\,{\rm pc\,cm}^{-3}$ signal above a signal-to-noise
ratio (SNR) of more than 5. Flagged time samples are replaced in the
filterbank file with noise drawn from adjacent unflagged samples, while
offending frequency channels are blanked. While the time-domain interference
excision weakens the SNR of bright signals of ${\rm DM} < 0.12
(w/64\,\mu$s)\,pc\,cm$^{-3}$, where $w$ is the pulse's width, it allows us to
produce and inspect candidates down to DM $\sim$1\,pc\,cm$^{-3}$ while
maintaining a manageable false detection rate at the manual inspection stage.

\begin{figure*}
\centering
\subfigure[]{
\includegraphics[width=0.7\textwidth,angle=270,trim=0mm 0mm 0mm 2mm, clip]{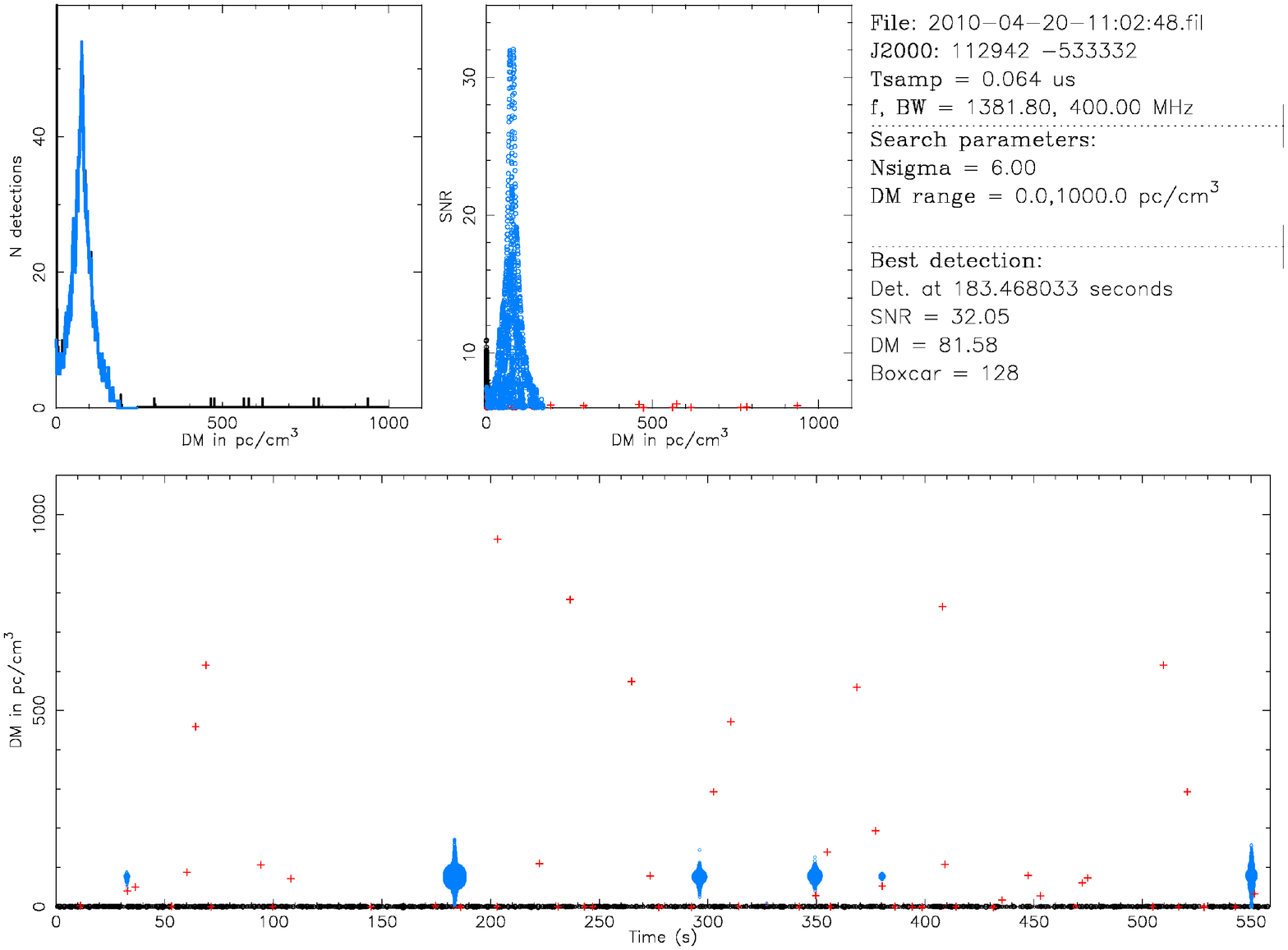}
}
\subfigure[]{
\includegraphics[width=0.43\textwidth,trim=0mm 0mm 0mm 0mm, clip]{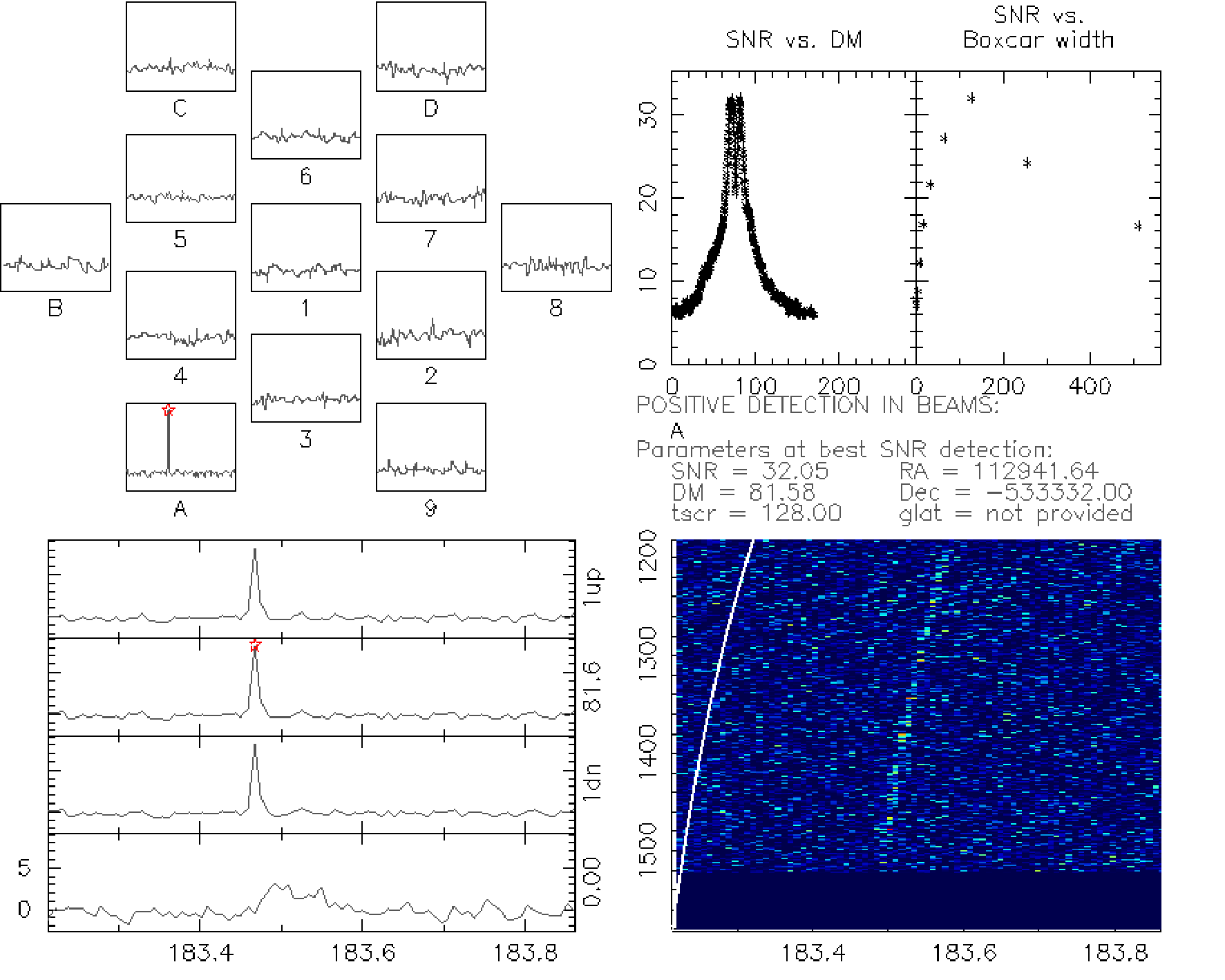}
}\quad
\subfigure[]{
\includegraphics[width=0.43\textwidth,trim=0mm 0mm 0mm 0mm, clip]{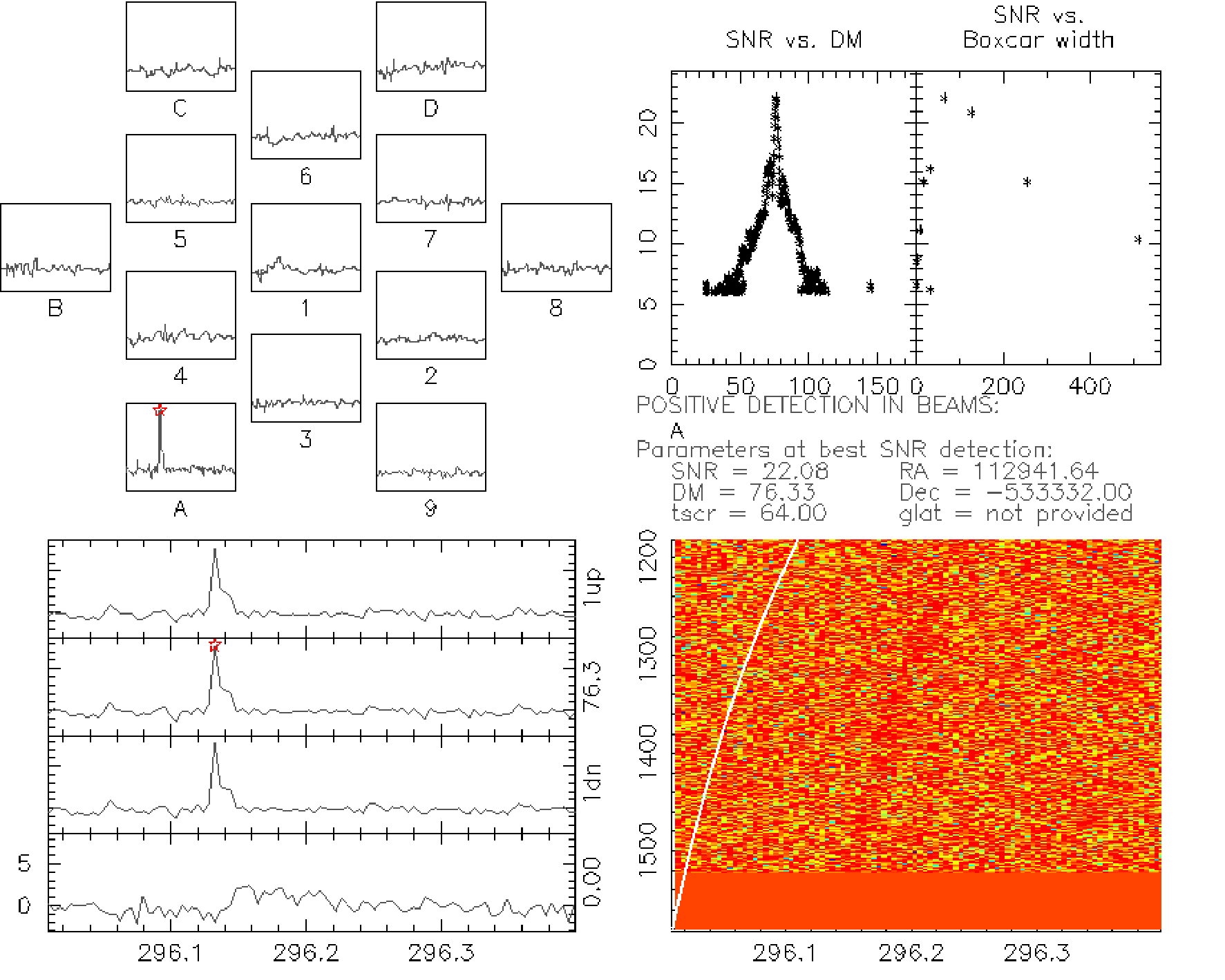}
}
\caption{Example diagrams used in the manual inspection stage. (a) Following
  \citet{cordesmclaughlin03}, a single-beam summary diagram showing a position
  near PSR J1129--53. 13 such plots are made for each pointing. Here events
  are coloured according to their auto-identified event type, as determined by
  the process described in \S\ref{sec:singlebeam}. Events flagged as zero-DM
  interference are black, those identified as gaussian peaks are red crosses,
  and candidate events are highlighted in blue/gray. The upper middle panel
  shows the SNR of each event plotted against DM, and the left panel shows a
  histogram of the net number of candidate (blue/gray) and all (black)
  detected event members in the pointing. The bottom panel shows the DM vs.
  time, with each member's SNR coded in the size of the plot point. In this
  example, six candidate pulses from PSR J1129--53 have been correctly
  discriminated from the noise and interference in the observation. (b,c) Here
  we show single-event plots that correspond to the second and third pulses
  detected from PSR J1129--53 in panel a. Panels show (clockwise from top
  left): the dedispersed timeseries in 13 beams; the SNR vs. DM (left
  subpanel) and boxcar filter trial (right subpanel) of the event; a
  false-colour image representing the signal power over frequency (MHz) and
  time since the beginning of the file (s); dedispersed time series at
  $DM=0\,{\rm pc\,cm^{-3}}$ (bottom subpanel), and three DM trial steps around
  the brightest detected DM (upwards from bottom subpanel). The structural
  deviations in the SNR vs.~DM curves in (b) and (c) from the predicted
  \citet{cordesmclaughlin03} curves are the combined result of noise, and both
  time- and frequency-dependent pulse structure.  }\label{fig:plots}
\end{figure*}

\begin{table}
  \begin{centering}
    \caption[HTRU survey observing parameters and status]{HTRU
      survey observing parameters and an indication of the processed portion
      of the survey. Full survey parameters are given in HTRU Paper
      1. Notes: $^\dag$The typical single-beam field of view is
      calculated within the average half-maximum beam width
      \mbox{($\simeq 0.044$\,deg$^2$).}}\label{table:parameters}
  \begin{tabular}{lll}
\hline
  & \multicolumn{2}{l}{ \textbf{Latitude region}} \\
\textbf{Parameter}& {\bf Inter.} & {\bf High} \\
 \hline
 \hline
Avg. pointing duration (s) & 540 & 270 \\
Total time processed  & 3490\,h & 140\,h \\
Total area processed$^{\dag}$ & 7671\,deg$^2$ & 591\,deg$^2$ \\
Fraction of total survey & 23.5\% & 0.39\% \\ 
\hline
Center frequency (MHz) & \multicolumn{2}{c}{1352} \\
Total bandwidth ($B$, MHz) & \multicolumn{2}{c}{340} \\
Typical usable $N_{\rm chan}$ & \multicolumn{2}{c}{870} \\
Sampling time ($t_{\rm samp}$, $\mu$s) & \multicolumn{2}{c}{64} \\
 \hline
\end{tabular}
\end{centering}
\end{table}

We perform dedispersion at 1196 trial DMs over the range 0 to $1000$
${\rm pc\,cm}^{-3}$ using the {\sc dedisperse\_all}
program.\footnote{This software and the C++ {\sc gtools} library,
  which is drawn from by {\sc dedisperse\_all} for single-pulse
  functionality and contains the SP search algorithms, candidate
  matching functions, and SP candidate type classes, are available
  from http://www.github.com/swinlegion}
After reading the input filterbank data into computer memory, time series are
formed for each DM trial; the efficient dedispersion aspect of the {\sc
  dedisperse\_all} code is described in detail by HTRU Paper 1. As the time
series data stream at each DM trial is produced, it is searched for single
pulses with parameters as in BSB. Our input search parameters differ only in
that the boxcar filter used in our search ranges in size $N_b = 1$ to 512
samples, and that events separated by more than 10 samples (640\,$\mu$s) were
considered independent. All significant (as defined by our pre-set SNR
threshold, $m_t>6$, detailed below) events in each boxcar and DM trial are
recorded and stored in a database. At the completion of dedispersion of all
trial DMs, {\sc dedisperse\_all} performs a temporal coincidence matching
(similar to the ``friends-of-friends'' method developed by \citet{FOF} and
used by \citet{deneva} and BSB to identify time-coincident event clusters).
Each event is defined by the parameters (DM, boxcar filter, width, time) at
which the signal-to-noise is found to be the greatest, and consists of members
at other DM and boxcar trials found to be coincident.

The presence of man-made interference (i.\,e.~radar communications, satellite
and aircraft transmissions, on-site hardware, and a number of other sources)
during observations causes the noise in our time series data to follow a
non-Gaussian distribution. These signals are typically either not dispersed,
or dispersed to a level that is undetectable in our data, and therefore the
zero-DM timeseries interference mitigation performed before processing removes
the bulk of these signals. After interference pre-removal, however, some
low-level inteference signals remain,
and we employ pre-set search thresholds to balance the false detection rate.
For most data, the interference pre-removal allows time series at trials above
${\rm DM} = 1.5$\,pc\,cm$^{-3}$, in the absence of our target astrophysical
signals, to be roughly Gaussian-distributed. At lower dispersion trials, the
false detection rate becomes unmanageable, and we therefore set a
dispersion-based threshold to reject all events found with a peak SNR at ${\rm
  DM} < 1.5$\,pc\,cm$^{-3}$. For each 9-minute intermediate latitude beam, we
produce 1196 dedispersed time series of $N_{s}\simeq8450000$\,samples. Our
total number of searched data points for each file consists therefore of
$1196\times\sum_{i=0}^{9}(N_s/2^i) \simeq 2.02\times10^{10}$. In
Gaussian-distributed data, at our SNR threshold of $m_t=6$ we would expect
approximately 50 independent random noise detections per file. We attempt to
filter these events by
automatically rejecting events which have less than three associated members.
This is an effective method of Gaussian event rejection; noise peaks at a SNR
of 12 or below are not likely to exhibit ${\rm SNR}>6$ in more than two
trials, assuming the peak's signal between DM steps and boxcar trials drops by
at least $\sqrt{2}$. Furthermore, the number of statistically random events
above ${\rm SNR}=12$ is negligible, even when considering analysis of the full
HTRU intermediate latitude survey. This strategy impacts weakly on our
sensitivity to events of $w=N_bt_{\rm samp}$ that have ${\rm SNR}\lesssim 12$.
However, in concert this event-match-based candidate rejection typically
reduces the number of candidates by a further factor of $\sim$100--1000, and
much more for data badly affected by interference. In Fig.\,\ref{fig:plots},
we show the two visual inspection plots used for manual candidate
descrimination. For the observation displayed, the telescope was pointed at
PSR J1129--53, a known RRAT (BSB). The six pulses emitted by the pulsar in
this pointing were correctly discriminated by the software from
zero-dispersion interference and from spurious peaks in the data.

All non-interference events are then written to disk by {\sc dedisperse\_all},
producing one ASCII file per event, in which all members of that event are
listed. At this stage, the single-pulse events for the beam are gathered by
the multibeam data monitor associated with the beam's pointing. The time
series are written to disk, stored for use in the periodicity search and for
later access by the multibeam data monitor.

\subsubsection{Multibeam Data Monitor}
As single-pulse events are produced for all the beams associated with one
pointing, the multibeam monitor for that pointing collects the events and
performs a temporal coincidence match of identical form to that done by {\sc
dedisperse\_all}, however it is performed for events from different beams.

The Parkes 20-cm multibeam receiving system allows the simultaneous observation
of 13 positions on the sky with approximately 30 arcminutes between beam
positions, and each beam has a sensitivity fall-off on scales of $<$30
arcminutes from its pointing centre \citep{multibeam}.  Pointlike radio
sources boresight to the telescope pointing direction will therefore typically
be detected in a maximum of three beams of the receiver.  Particularly
luminous objects like the Vela Pulsar, which emits single pulses with peak
flux densities of up to $\sim$60\,Jy, may be detected in up to $\sim$7 beams
when ideally positioned in the multibeam field. Typically, signals of
sufficient brightness to appear in all beams at similar intensity, i.\,e.
through a far sidelobe of the telescope, are generated by terrestrial or
near-Earth sources (such as satellites or aircraft). Therefore, as in BSB, we
do not inspect candidates which were detected in more than nine of the
thirteen beams.
This filter decreases our sensitivity to the terrestrial ``Perytons'' of
\citet{perytons}, however there is significant benefit gained through decrease
in the false detection rate per pointing. A search of the data aimed
specifically at detecting Perytons is planned for the near future. All events
occurring in nine beams or less are imaged as described in BSB, collecting the
relevant filterbank and time series data from the original on-disk location of
the single beams. When both single-pulse and pulsar searches have been
completed, the data is freed for removal from disk.

Finally, a pointing's result plots (i.\,e.~those for individual beams and
events as in Fig.\,\ref{fig:plots}) are manually assessed to determine whether
they contain a detection of interest. The ``beam summary'' plot has superiour
sensitivity to objects emitting multiple pulses with signals at or just
exceeding the detection threshold, while the single-event plots allow a user
swift discrimination between interesting candidates and falsely-identified
interference or noise. During the manual inspection stage, results are also
compared with the most up-to-date version of the Australia Telescope National
Facility Pulsar Catalogue originally published by \citet{psrcat}.

\subsection{Search Sensitivity} \label{sec:htrusens}
The sky coverage of the HTRU survey includes regions covered by previous
surveys that have been searched for single-pulses. Particularly for the HTRU
intermediate latitude survey, we have direct overlap with the areas searched
by BSB and PKSMB. Here we calculate our sensitivity to transient events and
make a comparison to these surveys.

\begin{figure}
\centering
\includegraphics[height=0.48\textwidth,trim=6mm 21mm 10mm 0mm, clip, angle=270]{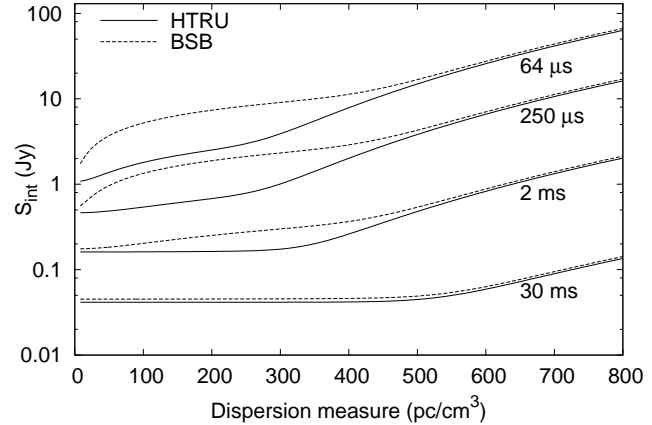}
\caption[]{{\small The HTRU survey's single-pulse search sensitivity to pulses
    of various intrinsic width, compared to the performance of the Parkes 20-cm
    analogue filterbank used by BSB (note: PKSMB had $t_{\rm samp}=250\,\mu$s
    whereas BSB had $t_{\rm samp}=125\,\mu$s; thus the PKSMB curve lies
    slighly above BSB's).  BPSR's higher frequency resolution most markedly
    improves our sensitivity to narrow pulses at low DMs; above ${\rm
    DM}\simeq$450\,pc\,cm$^{-3}$, the \citet{bhatetal04} model for
    interstellar scattering dominates the instrumental broadening for both
    surveys. Because of our limited search range in boxcar matched filter
    sizes, there is a sharp decrease in our sensitivity for pulses of duration
    $>32$\,ms. } }\label{fig:broadening}
\end{figure}

Instrumental and interstellar pulse broadening serves to weaken the SNR of the
pulse from its signal at the pulse's intrinsic width, $w_i$. All instrumental
broadening effects are in principle avoidable, therefore the back-end design
of any observing system aims to reach the scatter-broadening limit, as this is
the theoretical best that a pulsar/transients survey can achieve. As noted in
HTRU Paper 1, for a large range in DMs the improved time and frequency
resolutions of the HTRU survey causes pulse broadening from interstellar
propagation to dominate over intrachannel dispersion broadening ($t_{\rm
  ch}$). Using the empirical DM-scatter broadening relation given by
\citet{bhatetal04}, scatter broadening ($t_{\rm scat}$) typically dominates
over hardware-induced smearing above a DM of $\sim$230$\,{\rm pc\,cm}^{-3}$
for the HTRU survey (note, however, that the scatter of individual pulsars in
the Bhat et al.~relation means that individual objects may deviate by up to an
order of magnitude). To calculate our sensitivity to single pulses, we take
these pulse-scattering and broadening effects into account.

Following e.,g.~\citet{psrhandbook}, the observed pulse width depends on $w_i$
and various pulse-broadening effects by $w=(w_i^2+t_{\rm ch}^2+t_{\rm scat}^2
+ t_{\rm samp}^2)^{0.5}$. Note that our step size between DM trials is chosen
such that the broadening due to an error in DM is small compared to other
broadening effects. If the observed peak flux density for a pulse is given by
$S_{\rm peak}$, our sensitivity limit for the intrinsic peak flux density of
single pulses with observed duration $w=N_{\rm b}\times t_{\rm samp}$ is then
\begin{equation}\label{eq:sens}
	S_{\rm ilim} \geq S_{\rm peak}\cdot\frac{w}{w_i}~~;~~S_{\rm peak}=\frac{m_t\,T_{\rm sys}\,\beta}{G\sqrt{N_{\rm p}wB}}
\end{equation}
\citep[c.f.][]{psrhandbook}, where $T_{\rm sys}\simeq 23$\,K is the system
temperature of the multibeam system, $G$ is the telescope gain (ranging from
0.735--0.581\,K\,Jy$^{-1}$, from the central to outer beams), and $N_{\rm
  p}=2$ is the number of summed polarisations in the data. $\beta$ is a factor
of order $\sim$1 that is included to represent signal losses due to system
imperfections (for BPSR's two-bit digitisation, $\beta\simeq 1.07$,
following from \citealt{2bitstuff}). In Fig.\,\ref{fig:broadening}, we indicate
our sensitivity to single pulses of various durations as a function of DM, and
in comparison to the sensitivities of PKSMB and BSB.

The 2-bit digitisation levels of the BPSR instrument and large number of
frequency channels result in the increased dynamic range capabilities over the
previous Parkes analogue filterbank.
We made empirical measurements of the mean ($\mu$) and standard deviation
($\sigma$) of HTRU survey data judged by eye to not be strongly affected by
interference.
Using these values, we calculate the highest achievable SNR in the HTRU survey
data for pulses with $w=t_{\rm samp}$ to be $41.5$. This is in good agreement
with the SNR of saturated interference pulses observed in ${\rm DM}=0\,{\rm
pc\,cm}^{-3}$ time series. For the 1-bit, 96-channel analogue filterbank, the
maximum SNR may be calculated analytically, using a binomial probability
distribution.
The theoretical maximum SNR of this system at $w=t_{\rm samp}$ is then $9.6$;
again, this agrees with saturated interference signals observed in data from
this system. The differences in dynamic range for these observing backends do
not strictly impact the flux sensitivity of a single-pulse search, however the
HTRU survey's increase in dynamic range will affect the manual inspection
stage by allowing detections of potentially very bright, narrow pulses to be
more clearly discernible from noise or non-Gaussian statistics in the
observation.  This effect is more acute for pulses of width close to the
sampling time of each survey.

The improved time/frequency resolutions and dynamic range of the HTRU survey
is clearly beneficial for the detection of the narrow single pulses of pulsars
and RRATs, and extragalactic impulse emitters with durations $w\lesssim10$\,ms
and DMs of ${\rm DM}\lesssim 360\,{\rm pc\,cm}^{-3}$. For ${\rm DM}>360\,{\rm
  pc\,cm}^{-3}$, we do not expect to see a large number of new RRAT
discoveries for several reasons; first, we do not have a considerable increase
in sensitivity over the previous surveys for these DMs, and the number of
known RRATs above this DM represents only $\sim$10\% of the known population.
Additionally, for $|b|<5$, the 35-minute pointing duration of PKSMB searches
afford a greater probability of detection of low-pulsation rate events over
our 8.5-minute observations of these areas, and we offer minimal improvements
for the capability of detection of these objects. At higher latitudes, for the
bulk of pointing positions in the HTRU survey the number density of electrons
in the Galaxy is insufficient to produce a DM much greater than $360\,{\rm
  pc\,cm}^{-3}$ \citep{ne2001}. For these reasons, we expect most new rotating
radio transient and high nulling-fraction pulsar discoveries to be at DMs
${\rm DM}<360\,{\rm pc\,cm}^{-3}$. As detailed in \S\ref{sec:htrunewdets} and
\S\ref{sec:archcomp}, this is true for the initial discoveries.


\begin{figure*}
\centering \includegraphics[height=0.95\textheight,trim=1mm 2mm 1mm 1mm,
clip]{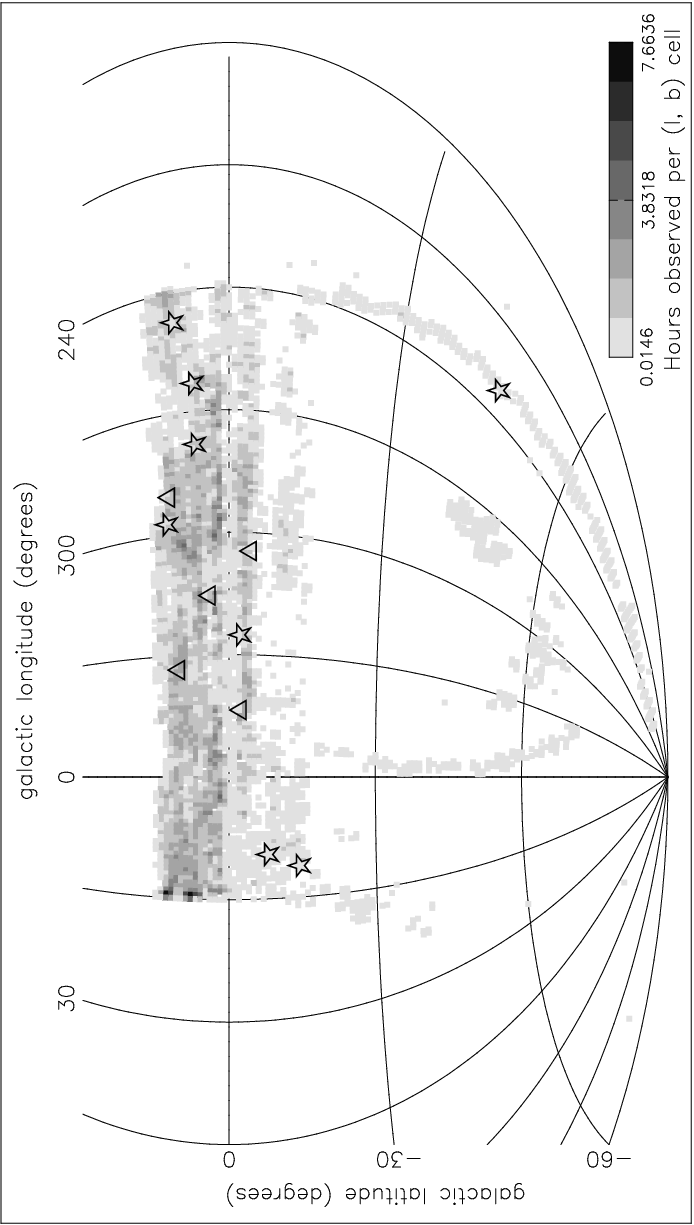}
\caption[]{A Galactic map showing new HTRU survey single-pulse discoveries,
  and observing time spent per $l, b$ cell. The stars indicate detections from
  which multiple pulses have been detected, and triangles indicate sources
  which have only yet exhibited one strong pulse.}\label{fig:galmap}
\end{figure*}

\subsection{Current Processing Status}
The current status of the HTRU survey's processing is summarised in the upper
panel of Table \ref{table:parameters}. The processing reported in this paper
includes only pointings from the HTRU intermediate and high latitude
surveys. The survey coverage to date is not contiguous over all Galactic
regions. We indicate the total observation time of the processed pointings as
a function of sky position in Figure \ref{fig:galmap}. In total, 22,273 and
1,718 beams have been processed out of the 95,056 and 443,287 total beams of
the intermediate and high latitude surveys, respectively. Approximately 160
incomplete survey pointings of shorter duration have also been processed. As
implied by Fig.\,\ref{fig:galmap}, the bulk of observing time reported in this
paper has been spent at Galactic latitudes $|b|<15$.


\subsection{Candidate Ranking and Follow-up Strategy}\label{sec:strategy}
We employ a follow-up prioritisation scheme for our single-pulse candidates to
provide maximum scientific return with available observing time. Based on the
strength of the candidate in various panels of the manual inspection plots and
the number of detections at a similar DM in a pointing
(Fig.\,\ref{fig:plots}), candidates are ranked as \emph{multi-pulse} (two or
more strong pulses detected at similar DM), \emph{strong} ($\geq$2 weak
pulses, or one convincing pulse detected, that is with a ${\rm SNR}>7$ and
exhibiting a characteristic signature in the SNR vs.~DM plot, e.\,g. as
explored by \citealt{cordesmclaughlin03}), or \emph{weak} (a $DM>0\,{\rm
pc\,cm^{-3}}$ detection appearing pointlike in the multibeam field, but poorly
supported by other inspection panels and/or has ${\rm SNR}<7$). \emph{Weak}
candidates outnumber those ranked \emph{strong} and \emph{multi-pulse} by a
factor of more than two. All candidates that are not found in available
archival data with pointings near the discovery position are reobserved for a
duration equal to that of the discovery pointing.
If \emph{weak} detections are not seen in the initial follow-up pointing, they
are not tracked in further follow-up observations. We do not include these
detections in this report due to their highly uncertain nature, however we
preserve them in a database for posterity, considering that borderline
detections in future surveys of the same sky area could cross-correlate their
findings with our weak, unconfirmed detections.
This will be possible with the HTRU low-latitude survey, and potentially with
future Australian Square Kilometre Array Pathfinder/Square Kilometre Array
surveys. \emph{Multi-pulse} and \emph{strong} candidates not seen in initial
follow-up are observed again for a duration of three times the discovery
pointing.
Our ranking levels are set such that the \emph{strong} and \emph{multi-pulse}
categories are highly unlikely to contain detections of spurious noise. We
include all such candidates here, with objects not detected in follow-up
observations distinguished in Table \ref{table:newdets} by an upper limit on
the objects pulsation rate ($\prate$, quoted as number of pulses per hour);
this reflects that non-detections of these discoveries are used only to place
limits on $\prate$ for the object. While the limits are well within the range
of pulsation rates for known RRATs, these objects \emph{essentially remain
  unconfirmed}, and care must be taken in including these objects in
statistical or categorical transients studies (note, however, that PSR
J1424--56 is confirmed by the reported detection of PSR J1423--56 by Keane
(2010b)\nocite{evanthesis}, as discussed below.

\begin{table*}
\begin{centering}
  \caption[]{Properties of the objects discovered in this search. Columns: (1)
    Name based on J2000 coordinate ($^\dag$ indicates a candidate from which
    only one clear pulse detection has yet been made. Such objects must be
    interpreted with care; see notes on these and the objects with limits only
    on $\prate$ in \S\ref{sec:strategy}). $^{*}$ PSRs J1307--67 and J1423--56
    are believed to be confirmations of PSRs J1308--67 and PSR J1424--56,
    respectively, of Keane (2010b)\nocite{evanthesis}; see notes about these
    object in Section \ref{sec:htrunewdets}; (2,3) J2000 right ascension and
    declination of pointing centre for detected beam. Only for J1854--1557,
    the position is as derived from the pulsar's timing solution; (4) the
    best-fit period, where measurable, with the error on the last digit in
    parentheses; (5) pulsation rate $\prate=N_{\rm p}$\,h$^{-1}$; (6) best-fit
    DM and error in pc\,cm$^{-3}$; (7) observed pulse width at half-maximum of
    the brightest pulse; (8) peak flux density of the brightest detected pulse
    (calculated using $S_{\rm peak}$ in Eq.\,\ref{eq:sens}).
  }\label{table:newdets}
\begin{tabular}{ccccccccc}
\hline {\bf } & \multicolumn{2}{c}{\bf Pointing centre} &  & $\prate$ &  & {\bf $w_{\rm eff}$} &{\bf S$_{\rm peak}$}\\ 
{\bf PSRJ} & {\bf RAJ} & {\bf DECJ} & {\bf P (s)} & {\bf(h$^{-1}$)} & {\bf DM} & {\bf (ms)} &{\bf (mJy)}\\ 
\hline 
J0410--31         & 04:10:39 & --31:07:29 & 1.8785(2)&     107 &9.2(3)   & 18 & 470 \\ 
J0837--24         & 08:37:44 & --24:47:48 &       -- &       5 &142.8(5) & 1  & 420 \\ 
J0912--38         & 09:12:27 & --38:48:34 & 1.5262(1)&      32 & 73.3(5) & 6  & 190 \\ 
J1014--48         & 10:14:18 & --48:49:42 & 1.5088(2)&      16 & 87(7)   & 21 & 140 \\ 
J1135--49$^{\dag}$ & 11:35:56 & --49:25:31 &       -- &$\leq$1.3& 114(20) & 9  & 120 \\ 
J1216--50         & 12:16:20 & --50:27:01 & 6.355(9) &      13 & 110(20) & 9  & 130 \\ 
J1307--67$^{*}$   & 13:07:41 & --67:03:27 &3.65120(8)&      11 & 47(15)  & 32 & 70 \\ 
J1424--56$^{*}$   & 14:24:23 & --56:40:47 &       -- & $\leq$7 & 27(5)   & 22 & 110 \\ 
J1541--42$^{\dag}$ & 15:41:55 & --42:18:50 &       -- & $\leq$7 & 60(10)  & 4  & 150 \\ 
J1549--57         & 15:49:05 & --57:21:37 & 0.7375(3)&      73 &17.7(3.5)& 4  & 210 \\ 
J1709--43$^{\dag}$& 17:09:47 & --43:54:43 &       -- & $\leq$7 & 228(20) & 3  & 240 \\ 
J1854--1557       &18:54:53.6&-15:57:22(20)& 3.4532(1)&      25 & 160(25) & 65 & 50 \\ 
J1925--16         & 19:25:06 & --16:01:00 & 3.8858(2)&$\leq$6.5& 88(20)  & 10 & 160 \\ 
\end{tabular}
\end{centering}
\end{table*}

\section{Early Discoveries \& Detections}\label{sec:htrunewdets}
\subsection{New Discoveries}\label{sec:newdisc}
Table \ref{table:newdets} lists basic parameters of the 13 new and confirmed
objects discovered thus far using the HTRU survey's single-pulse pipeline.
There are several objects worthy of specific note.

PSR J0410--31, at Galactic coordinates $l=230.6^\circ, b=-46.7^\circ$, is the
only confirmed single-pulse discovery arising thus far from the high-latitude
survey. The DM of this pulsar is low, with ${\rm DM}=9.2\,{\rm pc\,cm}^{-3}$,
lower than 98.9\% of known radio pulsars. Estimating distance to the new
discoveries using their DMs and the \citet{ne2001} electron density model for
the Galaxy (hereafter NE2001), we find that this is also the most nearby
object discovered in our search ($d=0.51$\,kpc).

PSR J0912---38 was also detected in the HTRU survey's Fourier-domain
search. Bright single pulses were only detectable above the noise for
$\lesssim$30 seconds in both the discovery and follow-up observations. When
the data are folded over pulses not detected by the single-event search, the
pulsar remains detectable with a signal-to-noise ratio of $>$5; this behaviour
is not consistent with scintillation. The abrupt change in flux density and
the occurrence of these short-duration flares in both observations lead us to
suggest they are intrinsic to the star.

The pulses of PSR J1014--48 were only detected in one cluster spread over 16
rotations of the neutron star. Like PSR J1825--33 of BSB, no outbursts have
yet been detected in follow-up observations.


\begin{figure}
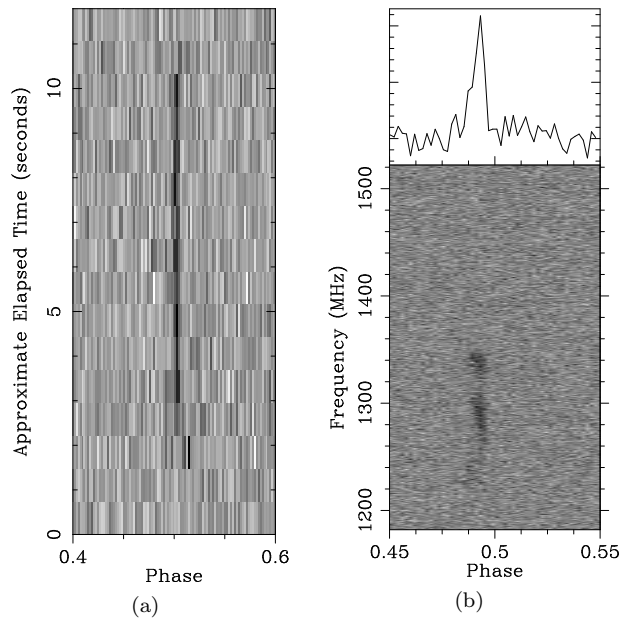

\centering
\subfigure[]{
\includegraphics[angle=270,width=0.21\textwidth,trim=2.5mm 0mm -3mm 0mm,clip]{figures/J1549a.ps}
}\quad
\subfigure[]{
\includegraphics[angle=270,width=0.22\textwidth,trim=0mm 0mm -3mm 0mm,clip]{figures/J1549b.ps}
}
\caption[]{PSR J1549--57: (a) A pulse stack of 16 single pulses. (b) A
  dedispersed waterfall image (lower panel) with dedispersed and
  band-integrated pulse profile (upper panel); these result after folding only
  over the 16 stellar rotations shown in (a).}\label{fig:1549}
\end{figure}

\begin{figure*}
\centering
\includegraphics[angle=180,width=0.75\textwidth]{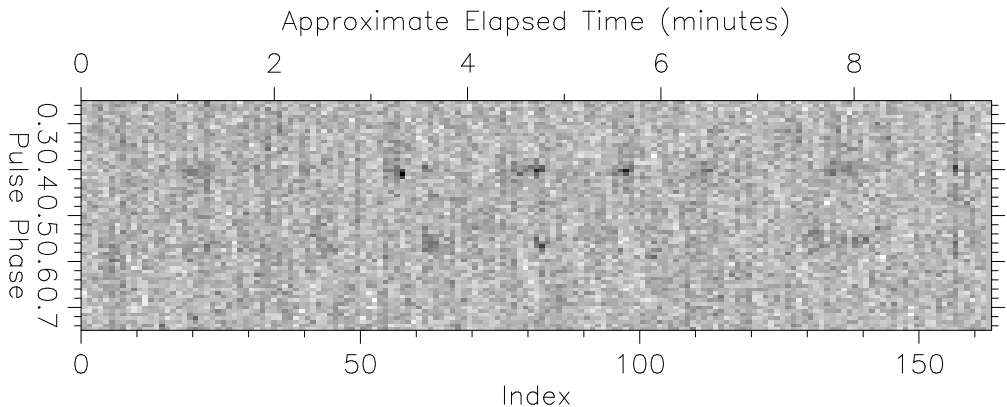}
\caption[]{A pulse stack showing single pulses of the periodically nulling
  pulsar J1854--1557.}\label{fig:1855}
\end{figure*}

We are confident that PSR J1307--67 is the PSR J1308--67 of Keane
(2010b)\nocite{evanthesis}, who reports a solitary pulse discovery at ${\rm
DM}=44\pm2{\rm pc\,cm^{-3}}$ and a position with errors in agreement with our
pointing position. Upon searching the archival data closest in position to our
discovery, we found 7 pulses with a signal-to-noise ratio of $>$5, from which
the period reported in Table \ref{table:newdets} was determined. Similarly, we
are confident that PSR J1424--56 is the PSR J1423--56 of Keane
(2010b)\nocite{evanthesis}, who reports ${\rm DM}=32.9\pm1.1{\rm pc\,cm^{-3}}$
and $\prate=3.4\,{\rm h}^{-1}$, both consistent with our findings. Timing
solutions will solve the positional and naming ambiguities of these objects.

Broad-bandwidth amplitude modulation features are seen in the
frequency-dependent structure of the single pulses of PSR J1549--57 (see
Fig.\,\ref{fig:1549}), the bandwidth of which are inconsistent with the
predicted scintillation bandwidth of NE2001 by four orders of magnitude. The
origin of this feature is unknown, however despite the significant
disagreement with the NE2001 prediction, it is likely to be scintillation if
the detected emission is produced by a neutron star.

The predicted scatter broadening of PSR J1709--43 (130\,ms at 1.375\,GHz for
the NE2001 model) is much greater than the half-maximum pulse width detected
here. The single pulse from this object had a signal to noise ratio of 10, a
visible dispersion curve, a signal-to-noise versus DM curve well-fit to the
model of \citet{cordesmclaughlin03} for a genuinely dispersed pulse, and
exists in a relatively interference-free observation. These conflicting
properties indicate that either the impulse's origin is not celestial, or that
the NE2001 scattering model for this Galactic position is incorrect; the
latter is highly possible to be the case here, as scattering is highly
dependent on line-of-sight. A redetection of this candidate would conclusively
clarify the nature of this candidate, however follow-up has not yet been
performed.

PSR J1854--1557 appears to be exhibiting periodic nulling, drifting, and
mode-changing behaviours (Fig.\,\ref{fig:1855}). Despite a high nulling
fraction, this object's relatively frequent emission renders it detectable in
a Fourier search with a signal to noise ratio of $\sim$17 in the 8.5-minute
survey and follow-up pointings. This object was also detected in the Fourier
pipeline, as reported in HTRU Paper 1.


\subsection{Redetections of Known Pulsars and RRATs}\label{sec:redets}
The single-pulse pipeline has detected approximately 55\% of the known pulsars
that could be detected through Fourier searches of the HTRU survey data. This
is an unexpectedly large fraction, and it offers improvements over the
single-pulse detection rates presented in BSB and PKSMB, perhaps a testament
to the increased sensitivity to faint and narrow single pulses that this
survey's hardware affords. The single-pulse properties of these objects will
be analysed in future work, so we will not detail them here.

We have processed 24 observations with pointing positions within 6 arcminutes
(our approximate half-power beam width) of published RRAT positions
(i.\,e.~from \citealt{rrats}, \citealt{mauranewpaper}, \citealt{keane},
\citealt{sbsmb}). Of these 24, 13 RRATs were redetected. For 22 of 24
observations, the number of pulses seen from each object was consistent with
the expected number based on previously published pulsation rates in the
discovery observations of these objects (again, as drawn from the BSB and
PKSMB publications) and the observation length in our data. The two notably
different detection rates were of RRATs J1753--38 (BSB; 4 expected and $>15$
seen) and J1819--1458 (\citealt{rrats}; 3 expected and 10 seen). PSR
J1819--1458 is known to exhibit small variations in pulsation rate \citep[and
larger ones associated with glitches,][]{lyne1819}, and the
detection rate observed here is consistent with the range in $\prate$ reported
by \citet{lyne1819}. For PSR J1753--38, the higher detection rate,
detectability in a Fourier search with a signal-to-noise ratio of $\sim$13
(whereas it was undetectable in a Fourier search of its original discovery
data), and the slight increase in $S_{\rm max}$ exhibited in our observations
indicate that: 1) this pulsar's emission appears to simply be highly modulated
rather than nulling, and 2) our observing position of the source is improved
from that reported by BSB.  Our positional offset was 0.1 degrees, suggesting
an improved position of right ascension and declination 17:53:09, --38:52:13
(J2000).


\section{Discussion}\label{sec:htrudiscussion}
\subsection{Our Discoveries in Archival Pulsar Surveys}\label{sec:archcomp}
All of our discoveries except for PSR J0410--31 and PSR J0837--24 lie in
regions previously covered by the surveys of \citet{BJ,ED} and \citet{pksmb}.
The presence of pulses in archival data can add valuable information about
pulse rate changes over time, and potentially add data points to their timing
if a phase-coherent timing solution is obtained. The detection of J1307--67
and J1424--56 in archival data have already been discussed in
Sec.~\ref{sec:newdisc}. We inspected data in the archival surveys within one
half-power beam width of our other objects' positions to determine the
detectability of our discoveries in the archival data.
None of our objects were detected in these pointings in single pulses.
However, while PSR J1854--1557 is not detectable in a Fourier search, it is
marginally detectable when the archival data are folded over the period given
in Table \ref{table:newdets}.


We investigate the reason for the non-detection of the remaining objects by
first estimating the signal loss due to the use of the analogue multibeam
filterbank system used by BSB and PKSMB, calculating the SNR that a pulse of
equivalent width, DM, and $S_{\rm peak}$ (as listed in Table
\ref{table:newdets})
would have shown in the archival data. This analysis suggests that even the
brightest detected pulses in our data would result in only a marginal (${\rm
  SNR}<6$) detection in the analogue filterbank data for PSRs J1135--49,
J1541--42, and J1709--43. For two out of four of the remaining objects at
$|b|>5^\circ$, for which the archival data length was 4.4\,minutes, the
non-detections are accounted for by the low probability that a pulse would
occur during the time span of the observation.
The remaining $|b|>5^\circ$ objects, PSR J0912--38 and PSR J1014--48, had
clusters of sequential pulses rather than a smooth distribution of single
pulses. Although the per-hour pulsation rate is high for these objects, the
duration of on-activity is short, and the spacing between sequential pulse
outbursts is $\gtrsim$8 minutes. This indicates that the non-detection of
pulses is consistent with a decreased probability of the occurrence of a pulse
outburst during the archival observations.
The effect of on- and off-timescales on object detectability will be discussed
further in Sec.~\ref{sec:clumping}.

The only remaining undetected object is PSR J1549--57. We have not yet
performed follow-up observations on this object, and so the archival data, in
which we would have expected to see detections of ${\rm SNR}>7$, serves only
to place limits on the pulsation rate and duration of off-activity in these
objects. The non-detection in two archival pointings places a strong limit
(zero pulses in $\geq$70\,m) on these values.

\newcommand{\tobs}{t_{\rm obs}} \newcommand{\ton}{t_{\rm on}}
\newcommand{\toff}{t_{\rm off}}
\subsection{Activity Timescales in Sparsely-emitting Neutron Stars}\label{sec:clumping}

Approximately 1/3 of our single-pulse pipeline discoveries exhibit distinct
periods of on-activity, marked by bright sequential pulses that are separated
by longer intervals of either genuine nulls, or a decrease in intensity level
sufficiently large that the object is undetectable through single pulse or
periodicity search techniques during the null state.
The pulsation rate, $\prate$, that is typically quoted for single-pulse
discoveries poorly represents the emissivity of such objects. It provides
insufficient information to assess either the probability of discovering such
a neutron star in a given observation, or of detecting such an object during a
reobservation.

Considering this and the significant capability of the HTRU survey to discover
new genuinely sparsely-emitting objects (see \S\ref{sec:forecast}), we seek to
more accurately incorporate our discoveries and RRAT populations in general
into the range of nulling and emissivity timescales exhibited by radio
pulsars. We explore a relevant ``intermittence parameter space'' here, and
review in \S\ref{sec:detectability} what effect survey parameters have on
the selection of various populations in this space.

We
note that the on- and off-states of nulling pulsars typically show
characteristic lengths \citep{herfindalrankin07,redmanrankin09,wangetal07},
parameterised below by $\ton$ and $\toff$, respectively.  This description of
neutron star intermittence is more physically representative for pulsars of
various nulling fractions, better reflects the timescales associated with
possible windowing phenomena, and allows a more accurate exploration of
selection effects for surveys of various length (see
\S\ref{sec:detectability}).  Here we use a definition of $\toff$ that is
similar to the ``null length'' of \citet{wangetal07}. We define $\toff$ as the
average time between the first pulse whose signal drops below a set threshold
(we use ${\rm SNR}>5$) and the next above that threshold. We define $\ton$ as
the average number of pulses ($N_{\rm on}$) above the same threshold, times
the rotational period of the object. Obviously, such values are only valid for
genuine nulling pulsars, that is this analysis is inappropriate for highly
modulated pulsars with long-tailed pulse energy distributions
\citep[e.\,g.][]{modpulsar}. Uninterrupted data spans of $T\gg\ton+\toff$
containing single-pulse detections of high significance (that is, where the
single pulse energy distribution of the on-pulses is such that all pulses
exceed the threshold) would provide the most accurate measurements of these
quantities; for our discoveries, most observations provide robust single-pulse
detections, although we are insensitive to timescales $\gg9$\,minutes. We
nevertheless make estimates of (or place limits on, in observations where no
timescales are measureable because they appear greater than our data span) the
average $\ton, \toff$ for our discoveries based on all available data.

The nulling fraction of a pulsar is only valid in this analysis when measured
over a timescale far exceeding one \emph{activity cycle} of the object, which
is given by a length $T_{\rm A} = \ton+\toff$. Below we generically represent
the nulling fraction measured over ``infinite time'' as $f_\infty$, where
accordingly, the pulsar's on fraction measured over a time $T\gg T_{\rm A}$ is
given by $\xi=1-f_\infty$.
As an example, the pulsar PSR B1931+24 exhibits periods of activity lasting
$\sim$5-7 days, separated by nulls of $\sim$30 days \citep{sometimesapulsar}.
While the nulling fraction measured over short observational timescales is
therefore typically either 0.0 or 1.0, we find that $f_\infty\simeq0.85$.

\begin{figure*}
\centering
\includegraphics[angle=270,width=0.95\textwidth,trim=5mm 12mm 0mm 0mm, clip]{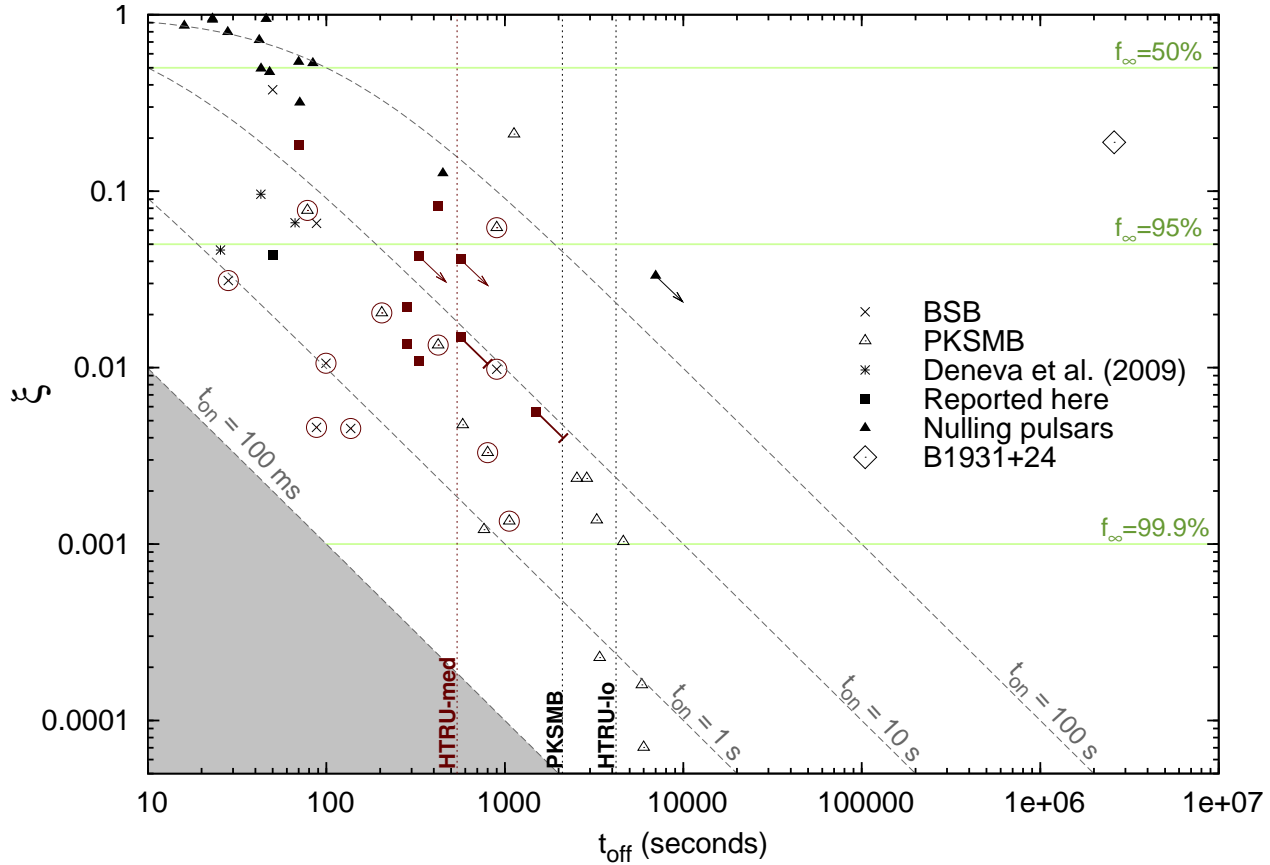}
\caption[On- and off-timescales for nulling pulsars]{Here we show the on
  fraction, $\xi$, against the mean null length for single-pulse search
  discoveries and the pulsars of \citet{wangetal07}. All elements highlighted
  in maroon indicate relevance to the HTRU intermediate latitude survey; the
  squares show single-pulse pipeline discoveries, while maroon circles
  indicate previously-known objects which were redetected in the HTRU
  intermediate latitude survey data (\S\ref{sec:redets}). Points with arrows
  designate an object for which $\ton$ has been measured but only a lower
  limit on the null timescale is known, and points with flat-headed arrows
  represent objects for which only one pulse has been detected (we have
  assumed $P=8.5$\,s for these objects, therefore the plotted point is an
  upper limit to $\xi$ and a lower limit to $\toff$). The solid green lines
  indicate lines of constant nulling fraction as measured over infinite time
  ($f_\infty$, see \S\ref{sec:clumping}), and the dashed gray lines show
  constant contours of $\ton$. The vertical dotted lines show the
  single-pointing observation length for the survey of PKSMB, and for the HTRU
  high (hi), intermediate (med), and low (lo) latitude surveys (see
  \S\ref{sec:detectability}). The shaded region below $\ton=100\,$ms indicates
  the area for which $\ton$ is shorter than the period range of single-pulse
  search discoveries, and is therefore ill-defined. B1931+24 is the
  ``intermittent pulsar'' published by \citet{sometimesapulsar}, and discussed
  in Sec.~\ref{sec:detectability}. }\label{fig:onoff}
\end{figure*}

In Figure \ref{fig:onoff}, we plot $\xi$ as a function of $\toff$ for the
single-pulse discoveries from HTRU, PKSMB, BSB, and \citet{deneva}. We also
plot all nulling pulsars analysed by \citet{wangetal07} that have a non-zero
nulling fraction. The Wang et al.~sample was selected from the
periodicity-discovered pulsars of the same survey as the PKSMB searches, and
has poor sensitivity to nulls of $\ll30$\,s due to the averaging performed
over 10-30\,s timescales. The dashed grey lines in this figure show contours
of constant $\ton$. For a pulsar of period $P$, it is not straight-forward to
identify on- or off-activity timescales of duration $t<P$; the shaded region
below $\ton=100\,$ms indicates the area for which $\ton$ is shorter than the
period range of single-pulse search discoveries that emit at $N_{\rm on}=1$.

One critical point illustrated by this plot is the continuous transition
between the \citet{wangetal07} nulling pulsars and the population discovered
by single-pulse searches. \citet{mclaughlincordes03} derive that neutron stars
of period $P$ will be more readily discovered in a single-pulse search,
(i.\,e.~having SNR ratio $r=m_{\rm SP}/m_{\rm FFT}>1$) when
\mbox{$\xi<2\sqrt{P/T}$}, leading Keane (2010a)\nocite{keaneproceeding} to
note that this effect causes $\xi$- and $P$-based selection effects for single
pulse searches.
For the HTRU intermediate latitude survey pointings, pulsars with periods
$P<6$\,s should therefore be discovered more readily in the Fourier pipeline
at $\xi>0.2$, and thus it is not surprising that no HTRU intermediate latitude
survey points lie above this line in Fig.~\ref{fig:onoff}.
For the $T=35$\,min pointings of the Parkes Multibeam survey \citep{pksmb},
$r>1$ when $\xi<0.02$~to~$0.11$, considering a period range 200\,ms to 6\,s.
Appropriately, this appears to be the region in Fig.~\ref{fig:onoff} where the
nulling pulsars give way to the single-pulse discoveries.
It is clear here that the on and off timescales, and the nulling fractions of
the radio pulsars discovered in single-pulse searches exhibit a smooth
distribution over many orders of magnitude with no obvious distinction of the
RRAT population represented by the single-pulse discoveries.
A rigorous nulling fraction analysis for bright pulsars in the HTRU
intermediate latitude survey is planned for the near future, and is capable of
addressing whether the objects discovered by single-pulse searches represent a
smooth extension in the probability density function of pulsar nulling
fractions.


\subsection{Detectability of Intermittent Neutron Star Populations}\label{sec:detectability}

The distribution of objects in Figure \ref{fig:onoff} is influenced by a
number of selection effects, which we explore here both to investigate the
underlying distribution of activity cycles in neutron stars, and to determine
the population to which the three HTRU survey components will be sensitive.

Considering the general detectability of pulsars in this phase space, the
probability that at least one pulse will be emitted by a bright pulsar during
an observation of length $T$ is given by:
\begin{equation}\label{eq:pemit}
P_{\rm em} = \begin{cases} 0 & {\rm for~} f_\infty = 1\\
 1 & {\rm for~} T>\toff\\
 (T-\xi T+\xi\toff)/\toff & {\rm for~} T\leq\toff
\end{cases}
\end{equation}
Note, however, that this probability is generally representative and will hold
for a sample of many sources, but will not always be strictly true for an
individual source as the distribution of $\ton$ and $\toff$ for a single
object is not a delta function. Following this equation, the vertical lines in
Fig.\,\ref{fig:onoff} mark the null length below which a sufficiently bright
emitter has a probability of unity that it will be detected in the Parkes
Multibeam survey, the HTRU intermediate latitude survey, and the (recently
commenced) low-latitude survey, respectively. At null lengths greater than the
observing time for these surveys, the probability of detection quickly
decreases.  Accordingly, given a population of pulsars with a flat underlying
distribution in $\toff$ or log($\toff$), we would expect to have a higher
discovery/detection rate of objects with $\toff<T$.  Focussing on the objects
detected by the HTRU intermediate latitude survey (highlighted in maroon in
Fig.\,\ref{fig:onoff}), the distribution is roughly consistent with a flat
distribution in log($\toff$), with a ratio of 7:13 for objects with
$\toff\geq9$\,min to objects with $\toff<9$\,min. We point out that the object
of the highest $\toff$ discovered in each of the PKSMB, BSB, and HTRU
intermediate latitude surveys is roughly a factor of three greater than the
corresponding survey's length.  This corresponds to $P_{\rm em}\gtrsim0.33$ at
all nulling fractions.

Equation \ref{eq:pemit} indicates that on average for $\toff<T$, the
probability a source would be detected is one, and for $\toff>T$, it is
directly proportional to $\xi$. It is apparent in Fig.~\ref{fig:onoff} that
there is a drop in Parkes Multibeam detections in the region $100\,{\rm
  s}<\toff<35$\,min, $\ton>100$\,s. That is, if the underlying neutron star
population has a flat distribution in both $\ton$ and $\toff$ or log($\ton$)
and log($\toff$), there appears to be a deficit of objects in this region.
For objects with $\xi\gtrsim0.1$ (that is, those theoretically detectable in a
Fourier-search), there is no explicitly stated reason that \citet{wangetal07}
should have selected against this range of timescales.  The paucity of objects
occupying this region could be the combined result of a number of possible
effects.  Nulling behaviour may have not yet been either observed or
recognised, however we would expect this to only be the case for objects with
$\ton\gtrsim35$\,min, which encompasses only a small fraction of the region in
question. It is possible that some low or intermediate nulling-fraction
objects with $\toff\simeq35$ were outright rejected from the survey due to a
non-appearance in follow-up observations. We consider this an unlikely cause,
however, as the 6-minute grid pointings and subsequent 35-minute follow-up
pointings (in case of a non-detection in the grid observation,
\citealt{pksmb}) would have provided a sufficiently high redetection
probability, particularly for objects with $\toff\ll T$. We note that for
objects with $\toff$ close to 35\,minutes, the survey's follow-up observations
might contain only a few pulses from the objects, which are detectable in
single pulses, rather than in the Fourier search used to assess the follow-up
pointings in the survey. However, this effect still does not account for the
notable detection rate drop in the region in question.

It is furthermore possible that the lack of objects in this region reflects an
actual drop in the neutron star population for objects with
$\ton\gtrsim300$\,s. This supposition is questioned, however, by the existence
of the ``intermittent pulsar'' B1931+24 reported by \citet{sometimesapulsar}.
This object is isolated in this space, although it has been indicated that a
handful of further unpublished objects occupy a similar region of
Fig.\,\ref{fig:onoff} \citep{mikethesis,ralphthesis}. The existence of these
objects suggests that either there is a bridging distribution of neutron stars
which we have not yet identified due to an unrecognised effect of selection,
or that in fact the intermittence of these objects is produced by a physically
distinct mechanism. Immense observing timescales are required for a survey to
have a high probability of detecting a population between the
$\ton\lesssim300$\,s objects and PSR B1931+24-like pulsars. Because data
storage and analysis becomes problematic for pulsar surveys of longer
timescale, we suggest that the brightest pulsars in this population may be
more readily uncovered by the ``transient-imaging'' surveys that are being
performed and planned on arrays; these surveys can sample timescales down to
the correlator averaging time (typically $\gtrsim$10\,s) and up to several
hours to days. However, for pulsars with $\ton$ less than these timescales,
pulsar surveys remain the most robust probes of intermittent populations.

\subsection{Discovery Forecast for the Full HTRU Survey}\label{sec:forecast}
The rate of new single-pulse discoveries in the HTRU survey, particularly
those in overlapping regions of previous surveys, hint at the potential for
the full HTRU survey to uncover many new examples of sparsely-emitting neutron
stars.  For the intermediate latitude survey, we have detected a total of 26
new and known single-pulse emitters, suggesting detection rates on the order
of $3\times10^{-3}$\,deg$^{-2}$ at these latitudes. We have processed roughly
23.5\% of the intermediate latitude survey to date, and if our discovery rate
continues, the full HTRU intermediate latitude survey data should contribute
an additional estimated $\sim$50 new discoveries of transient neutron stars.
We note that the sky distribution of neutron stars should have a higher
density at lower Galactic latitudes, and as seen in Figure \ref{fig:galmap} we
have not yet processed most of the pointings at latitudes closest to $b=0$. At
higher latitudes we have only processed a small fraction of the survey, and
while tempting to take an extrapolated value as an upper limit,
we note that our single discovery in only 0.39\% of the high latitude survey
implies a sky density of $\sim$$2\times10^{-3}$\,deg$^{-2}$, higher than
expected when compared to the sky density at lower latitudes. Therefore, the
detections are expected to be less than 250.


Finally, the 
explorations of Sections \ref{sec:clumping} and \ref{sec:detectability} have
accented the timescales of neutron star intermittence accessable by the HTRU
survey. In particular, the HTRU low-latitude survey (with its 70-minute
pointing length) will mark a significant increase in detection rate over the
PKSMB survey for objects with $35<\toff<70$\,min, and may be realistically
expected to discover objects with null lengths up to $\toff=3.5$\,h, if such
objects exist (as deduced in \S\ref{sec:detectability}, in which it was noted
that transients surveys have all been able to discover objects with $\toff$
values of up to 3 times the survey pointing length; see also
Fig.~\ref{fig:onoff}). The low latitude survey will also provide data that may
be used to investigate the possible deficit of low-nulling-fraction objects
with $\ton>300$\,s. Care must be taken during both the single-pulse and
periodicity searches, analysis, and follow-up, such that if these objects do
exist, they are not selected against. This will involve, for one, appropriate
follow-up monitoring of the stronger Fourier candidates that are undetected in
initial follow-up pointings.


\section{Summary and Conclusions}
We have presented the methods of and initial discoveries for the single-pulse
analysis of the High Time Resolution Universe Survey. We outlined the design
of the HTRU survey's single-pulse pipeline, which functions efficiently
alongside the Fourier-analysis pipeline; the single-pulse analysis furthermore
employs a ``friend of friends'' single-event recognition algorithm and
performs automated interference rejection based on the dispersive and
multi-beam signature of single events. The new digital backend used to collect
HTRU survey data has afforded a factor of up to $\sim$5 times improvement in
sensitivitiy over previous surveys in overlapping Galactic regions, and offers
the most significant improvements for short (sub-ms duration) pulses at low
(${\rm DM}<360\,{\rm pc\,cm}^{-3}$) DMs.

Analysis of 23.5\% and 0.39\% of HTRU intermediate and high Galactic latitude
survey data, respectively, has resulted in the discovery of 12 and 1 new
neutron stars. Much of the survey pointings covered Galactic regions that were
previously surveyed and searched for single pulses \citep{rrats,keane,sbsmb};
11 of our new discoveries lie within these regions, and for the nine of these
that were not visible in the archival data, their non-detection was consistent
with either the signal degradation due to the use of the previous
wider-channel analogue backend, or with the improbability of a pulse being
emitted in the archival survey due to a long null cycle.

Finally, we investigated the distribution of nulling and emissivity timescales
for the new single-pulse neutron star discoveries and redetections in the HTRU
survey data, and for the RRAT/nulling population in general. We found that
periodicity-discovered nulling pulsars and single-pulse search discoveries
exhibit a continuous distribution across null/activity timescales and nulling
fractions, building on evidence that many RRATs represent a tail of
extreme-nulling pulsars. We found that there is an apparent decrease in objects
with emissivity cycles longer than $\sim$300\,seconds at intermediate and low
nulling fractions which is not readily explained by selection effects, and
note that the HTRU deep low-latitude survey will be capable of exploring
whether this deficit is natural or an effect of selection. Lastly, we
estimated that the full HTRU survey is capable of more than doubling the known
extreme-nulling pulsar population, and will explore the neutron star
population with nulling fractions exceeding 99.99\%, and null lengths lasting
up to 3-4 hours.

\bibliographystyle{mn2e}
\bibliography{htruSP}

\begin{thebibliography}{}

\bibitem[\protect\citeauthoryear{{Bhat}, {Cordes}, {Camilo}, {Nice} \&
  {Lorimer}}{{Bhat} et~al.}{2004}]{bhatetal04}
{Bhat} N.~D.~R.,  {Cordes} J.~M.,  {Camilo} F.,  {Nice} D.~J.,    {Lorimer}
  D.~R.,  2004, \apj, 605, 759

\bibitem[\protect\citeauthoryear{{Burke-Spolaor} \& {Bailes}}{{Burke-Spolaor}
  \& {Bailes}}{2010}]{sbsmb}
{Burke-Spolaor} S.,  {Bailes} M.,  2010, \mnras, 402, 855

\bibitem[\protect\citeauthoryear{{Burke-Spolaor}, {Bailes}, {Ekers}, {Macquart}
  \& {Crawford} III}{{Burke-Spolaor} et~al.}{2011}]{perytons}
{Burke-Spolaor} S.,  {Bailes} M.,  {Ekers} R.,  {Macquart} J.,    {Crawford}
  III F.,  2011, \apj, 727, 18

\bibitem[\protect\citeauthoryear{{Cairns}, {Johnston} \& {Das}}{{Cairns}
  et~al.}{2001}]{cjd01}
{Cairns} I.~H.,  {Johnston} S.,    {Das} P.,  2001, \apjl, 563, L65

\bibitem[\protect\citeauthoryear{{Cordes} \& {Lazio}}{{Cordes} \&
  {Lazio}}{2002}]{ne2001}
{Cordes} J.~M.,  {Lazio} T.~J.~W.,  2002, ArXiv Astrophysics e-prints

\bibitem[\protect\citeauthoryear{{Cordes} \& {McLaughlin}}{{Cordes} \&
  {McLaughlin}}{2003}]{cordesmclaughlin03}
{Cordes} J.~M.,  {McLaughlin} M.~A.,  2003, \apj, 596, 1142

\bibitem[\protect\citeauthoryear{{Deneva}, {Cordes}, {McLaughlin}, {Nice},
  {Lorimer}, {Crawford}, {Bhat}, {Camilo}, {Champion}, {Freire} \& et
  al.}{{Deneva} et~al.}{2009}]{deneva}
{Deneva} J.~S.,  {Cordes} J.~M.,  {McLaughlin} M.~A.,  {Nice} D.~J.,  {Lorimer}
  D.~R.,  {Crawford} F.,  {Bhat} N.~D.~R.,  {Camilo} F.,  {Champion} D.~J.,
  {Freire} P.~C.~C.,    et al. 2009, \apj, 703, 2259

\bibitem[\protect\citeauthoryear{{Eatough}}{{Eatough}}{2009}]{ralphthesis}
{Eatough} R.~P.,  2009, PhD thesis, University of Manchester

\bibitem[\protect\citeauthoryear{{Edwards}, {Bailes}, {van Straten} \&
  {Britton}}{{Edwards} et~al.}{2001}]{ED}
{Edwards} R.~T.,  {Bailes} M.,  {van Straten} W.,    {Britton} M.~C.,  2001,
  \mnras, 326, 358

\bibitem[\protect\citeauthoryear{{Herfindal} \& {Rankin}}{{Herfindal} \&
  {Rankin}}{2007}]{herfindalrankin07}
{Herfindal} J.~L.,  {Rankin} J.~M.,  2007, \mnras, 380, 430

\bibitem[\protect\citeauthoryear{{Hessels}, {Ransom}, {Kaspi}, {Roberts},
  {Champion} \& {Stappers}}{{Hessels} et~al.}{2008}]{hesselsproc}
{Hessels} J.~W.~T.,  {Ransom} S.~M.,  {Kaspi} V.~M.,  {Roberts} M.~S.~E.,
  {Champion} D.~J.,    {Stappers} B.~W.,  2008, in {C.~Bassa, Z.~Wang,
  A.~Cumming, \& V.~M.~Kaspi} ed., 40 Years of Pulsars: Millisecond Pulsars,
  Magnetars and More Vol.~983 of American Institute of Physics Conference
  Series, {The GBT350 Survey of the Northern Galactic Plane for Radio Pulsars
  and Transients}.
pp 613--615

\bibitem[\protect\citeauthoryear{{Huchra} \& {Geller}}{{Huchra} \&
  {Geller}}{1982}]{FOF}
{Huchra} J.~P.,  {Geller} M.~J.,  1982, \apj, 257, 423

\bibitem[\protect\citeauthoryear{{Jacoby}, {Bailes}, {Ord}, {Edwards} \&
  {Kulkarni}}{{Jacoby} et~al.}{2009}]{BJ}
{Jacoby} B.~A.,  {Bailes} M.,  {Ord} S.~M.,  {Edwards} R.~T.,    {Kulkarni}
  S.~R.,  2009, \apj, 699, 2009

\bibitem[\protect\citeauthoryear{{Johnston} \& {Romani}}{{Johnston} \&
  {Romani}}{2003}]{johnstonromani03}
{Johnston} S.,  {Romani} R.~W.,  2003, \apjl, 590, L95

\bibitem[\protect\citeauthoryear{Keane}{Keane}{010a}]{keaneproceeding}
Keane E.~F.,  2010a, proceedings of ``High Time Resolution Astrophysics IV -
  The Era of Extremely Large Telescopes''

\bibitem[\protect\citeauthoryear{{Keane}}{{Keane}}{010b}]{evanthesis}
{Keane} E.~F.,  2010b, PhD thesis, University of Manchester

\bibitem[\protect\citeauthoryear{{Keane} \& {Kramer}}{{Keane} \&
  {Kramer}}{2008}]{evan}
{Keane} E.~F.,  {Kramer} M.,  2008, \mnras, 391, 2009

\bibitem[\protect\citeauthoryear{{Keane}, {Ludovici}, {Eatough}, {Kramer},
  {Lyne}, {McLaughlin} \& {Stappers}}{{Keane} et~al.}{2010}]{keane}
{Keane} E.~F.,  {Ludovici} D.~A.,  {Eatough} R.~P.,  {Kramer} M.,  {Lyne}
  A.~G.,  {McLaughlin} M.~A.,    {Stappers} B.~W.,  2010, \mnras, 401, 1057

\bibitem[\protect\citeauthoryear{{Keith}}{{Keith}}{2007}]{mikethesis}
{Keith} M.~J.,  2007, PhD thesis, University of Manchester

\bibitem[\protect\citeauthoryear{Keith, Jameson, {van Straten}, Bailes,
  Johnston, Kramer, Possenti, Bates, Bhat, Burgay, Burke-Spolaor, D'Amico,
  Levin, McMahon, Milia \& Stappers}{Keith et~al.}{2010}]{HTRU1}
Keith M.~J.,  Jameson A.,  {van Straten} W.,  Bailes M.,  Johnston S.,  Kramer
  M.,  Possenti A.,  Bates S.~D.,  Bhat N.~D.~R.,  Burgay M.,  Burke-Spolaor
  S.,  D'Amico N.,  Levin L.,  McMahon P.~L.,  Milia S.,    Stappers B.~W.,
  2010, \mnras, p.~1356

\bibitem[\protect\citeauthoryear{{Kouwenhoven} \& {Voute}}{{Kouwenhoven} \&
  {Voute}}{2001}]{2bitstuff}
{Kouwenhoven} M.~L.~A.,  {Voute} J.~L.~L.,  2001, A\&A, 378, 700

\bibitem[\protect\citeauthoryear{{Kramer}, {Lyne}, {O'Brien}, {Jordan} \&
  {Lorimer}}{{Kramer} et~al.}{2006}]{sometimesapulsar}
{Kramer} M.,  {Lyne} A.~G.,  {O'Brien} J.~T.,  {Jordan} C.~A.,    {Lorimer}
  D.~R.,  2006, Science, 312, 549

\bibitem[\protect\citeauthoryear{{Lorimer}, {Bailes}, {McLaughlin}, {Narkevic}
  \& {Crawford}}{{Lorimer} et~al.}{2007}]{LB}
{Lorimer} D.~R.,  {Bailes} M.,  {McLaughlin} M.~A.,  {Narkevic} D.~J.,
  {Crawford} F.,  2007, Science, 318, 777

\bibitem[\protect\citeauthoryear{{Lorimer} \& {Kramer}}{{Lorimer} \&
  {Kramer}}{2005}]{psrhandbook}
{Lorimer} D.~R.,  {Kramer} M.,  2005, {Handbook of Pulsar Astronomy}

\bibitem[\protect\citeauthoryear{{Lyne}, {McLaughlin}, {Keane}, {Kramer},
  {Espinoza}, {Stappers}, {Palliyaguru} \& {Miller}}{{Lyne}
  et~al.}{2009}]{lyne1819}
{Lyne} A.~G.,  {McLaughlin} M.~A.,  {Keane} E.~F.,  {Kramer} M.,  {Espinoza}
  C.~M.,  {Stappers} B.~W.,  {Palliyaguru} N.~T.,    {Miller} J.,  2009,
  \mnras, 400, 1439

\bibitem[\protect\citeauthoryear{{Manchester}, {Hobbs}, {Teoh} \&
  {Hobbs}}{{Manchester} et~al.}{2005}]{psrcat}
{Manchester} R.~N.,  {Hobbs} G.~B.,  {Teoh} A.,    {Hobbs} M.,  2005, \aj, 129,
  1993

\bibitem[\protect\citeauthoryear{{Manchester}, {Lyne}, {Camilo}, {Bell},
  {Kaspi}, {D'Amico}, {McKay}, {Crawford}, {Stairs}, {Possenti}, {Kramer} \&
  {Sheppard}}{{Manchester} et~al.}{2001}]{pksmb}
{Manchester} R.~N.,  {Lyne} A.~G.,  {Camilo} F.,  {Bell} J.~F.,  {Kaspi} V.~M.,
   {D'Amico} N.,  {McKay} N.~P.~F.,  {Crawford} F.,  {Stairs} I.~H.,
  {Possenti} A.,  {Kramer} M.,    {Sheppard} D.~C.,  2001, \mnras, 328, 17

\bibitem[\protect\citeauthoryear{{McLaughlin} \& {Cordes}}{{McLaughlin} \&
  {Cordes}}{2003}]{mclaughlincordes03}
{McLaughlin} M.~A.,  {Cordes} J.~M.,  2003, \apj, 596, 982

\bibitem[\protect\citeauthoryear{{McLaughlin}, {Lyne}, {Keane}, {Kramer},
  {Miller}, {Lorimer}, {Manchester}, {Camilo} \& {Stairs}}{{McLaughlin}
  et~al.}{2009}]{mauranewpaper}
{McLaughlin} M.~A.,  {Lyne} A.~G.,  {Keane} E.~F.,  {Kramer} M.,  {Miller}
  J.~J.,  {Lorimer} D.~R.,  {Manchester} R.~N.,  {Camilo} F.,    {Stairs}
  I.~H.,  2009, \mnras, 400, 1431

\bibitem[\protect\citeauthoryear{{McLaughlin}, {Lyne}, {Lorimer}, {Kramer},
  {Faulkner}, {Manchester}, {Cordes}, {Camilo}, {Possenti}, {Stairs} \& et
  al.}{{McLaughlin} et~al.}{2006}]{rrats}
{McLaughlin} M.~A.,  {Lyne} A.~G.,  {Lorimer} D.~R.,  {Kramer} M.,  {Faulkner}
  A.~J.,  {Manchester} R.~N.,  {Cordes} J.~M.,  {Camilo} F.,  {Possenti} A.,
  {Stairs} I.~H.,    et al. 2006, \nat, 439, 817

\bibitem[\protect\citeauthoryear{{Miller}, {McLaughlin}, {Keane}, {Kramer},
  {Lyne}, {Lorimer}, {Manchester} \& {Camilo}}{{Miller} et~al.}{tted}]{miller}
{Miller} J.~J.,  {McLaughlin} M.~A.,  {Keane} E.~F.,  {Kramer} M.,  {Lyne}
  A.~G.,  {Lorimer} D.~R.,  {Manchester} R.~N.,    {Camilo} F.,  submitted,
  \mnras

\bibitem[\protect\citeauthoryear{{Redman} \& {Rankin}}{{Redman} \&
  {Rankin}}{2009}]{redmanrankin09}
{Redman} S.~L.,  {Rankin} J.~M.,  2009, \mnras, 395, 1529

\bibitem[\protect\citeauthoryear{{Rubio-Herrera}}{{Rubio-Herrera}}{2010}]{herr%
erathesis}
{Rubio-Herrera} E.~A.,  2010, PhD thesis, Universiteit van Amsterdam

\bibitem[\protect\citeauthoryear{{Staveley-Smith}, {Wilson}, {Bird}, {Disney},
  {Ekers}, {Freeman}, {Haynes}, {Sinclair}, {Vaile}, {Webster} \&
  {Wright}}{{Staveley-Smith} et~al.}{1996}]{multibeam}
{Staveley-Smith} L.,  {Wilson} W.~E.,  {Bird} T.~S.,  {Disney} M.~J.,  {Ekers}
  R.~D.,  {Freeman} K.~C.,  {Haynes} R.~F.,  {Sinclair} M.~W.,  {Vaile} R.~A.,
  {Webster} R.~L.,    {Wright} A.~E.,  1996, PASA, 13, 243

\bibitem[\protect\citeauthoryear{Wang, Manchester \& Johnston}{Wang
  et~al.}{2007}]{wangetal07}
Wang N.,  Manchester R.~N.,    Johnston S.,  2007, \mnras, 377, 1383

\bibitem[\protect\citeauthoryear{{Weltevrede}, {Johnston} \&
  {Espinoza}}{{Weltevrede} et~al.}{2010}]{trinity}
{Weltevrede} P.,  {Johnston} S.,    {Espinoza} C.~M.,  2010, ArXiv e-prints

\bibitem[\protect\citeauthoryear{{Weltevrede}, {Stappers}, {Rankin} \&
  {Wright}}{{Weltevrede} et~al.}{2006}]{modpulsar}
{Weltevrede} P.,  {Stappers} B.~W.,  {Rankin} J.~M.,    {Wright} G.~A.~E.,
  2006, \apjl, 645, L149

\end{thebibliography}

\end{document}